\documentclass[journal,twoside,web]{ieeecolor}
\usepackage{generic}
\usepackage{cite}
\usepackage{amsmath,amssymb,amsfonts}
\usepackage{algorithmic}
\usepackage{graphicx}
\usepackage{algorithm,algorithmic}
\usepackage{hyperref}
\hypersetup{hidelinks=true}
\usepackage{textcomp}
\usepackage{booktabs}
\usepackage{makecell}
\newenvironment{renum}
  {\begin{enumerate}}
  {\end{enumerate}}
\usepackage{scalerel}
\DeclareMathOperator*{\concat}{\scalerel*{\Vert}{\sum}}

\def\BibTeX{{\rm B\kern-.05em{\sc i\kern-.025em b}\kern-.08em
    T\kern-.1667em\lower.7ex\hbox{E}\kern-.125emX}}
\begin{document}
\title{Multi-modality Regional Alignment Network for Covid X-Ray Survival Prediction and Report Generation}
\author{Zhusi Zhong, Jie Li, John Sollee, Scott Collins, Harrison Bai, Paul Zhang, Terrence Healey, Michael Atalay, Xinbo Gao, \IEEEmembership{Fellow, IEEE}, Zhicheng Jiao, \IEEEmembership{Member, IEEE}
\thanks{(Corresponding author: Zhicheng Jiao)}
\thanks{Zhusi Zhong is with School of Electronic Engineering, Xidian University, Xi’an 710071, China; Department of Diagnostic Imaging, Rhode Island Hospital, Providence, RI 02903 USA; Warren Alpert Medical School of Brown University, Providence, RI 02903 USA (e-mail: zhongzhusi@stu.xidian.edu.cn; zhusi\_zhong@brown.edu).}
\thanks{Jie Li and Xinbo Gao are with School of Electronic Engineering, Xidian University, Xi’an 710071, China (e-mail: leejie@mail.xidian.edu.cn; xbgao@mail.xidian.edu.cn).}
\thanks{John Sollee, Scott Collins, Terrence Healey, Michael Atalay and Zhicheng Jiao are with Department of Diagnostic Imaging, Rhode Island Hospital, Providence, RI 02903 USA, and also with Warren Alpert Medical School of Brown University, Providence, RI 02903 USA (e-mail: john\_sollee@brown.edu; scollins1@Lifespan.org; thealey@lifespan.org; Michael\_Atalay@brown.edu; zhicheng\_jiao@brown.edu).}
\thanks{Harrison Bai is with Department of Radiology and Radiological Sciences, Johns Hopkins University School of Medicine, Baltimore, MD 21205 USA (e-mail: hbai7@jhu.edu).}
\thanks{Paul Zhang, is with Department of Pathology and Laboratory Medicine, Hospital of the University of Pennsylvania, Philadelphia, PA 19104 USA (e-mail: Paul.Zhang2@pennmedicine.upenn.edu).}}
\maketitle

\begin{abstract}
In response to the worldwide COVID-19 pandemic, advanced automated technologies have emerged as valuable tools to aid healthcare professionals in managing an increased workload by improving radiology report generation and prognostic analysis. This study proposes Multi-modality Regional Alignment Network (MRANet), an explainable model for radiology report generation and survival prediction that focuses on high-risk regions. By learning spatial correlation in the detector, MRANet visually grounds region-specific descriptions, providing robust anatomical regions with a completion strategy. The visual features of each region are embedded using a novel survival attention mechanism, offering spatially and risk-aware features for sentence encoding while maintaining global coherence across tasks. A cross LLMs alignment is employed to enhance the image-to-text transfer process, resulting in sentences rich with clinical detail and improved explainability for radiologist. Multi-center experiments validate both MRANet's overall performance and each module's composition within the model, encouraging further advancements in radiology report generation research emphasizing clinical interpretation and trustworthiness in AI models applied to medical studies. The code is available at https://github.com/zzs95/MRANet.
\end{abstract}


\section{Introduction}
\label{sec:introduction}
The COVID-19 pandemic has created a critical and sustained worldwide public health challenge. Rapid viral transmission has led to widespread infection, depleting healthcare resources \cite{world2020announces}. Prompt diagnosis and clinical intervention is imperative for mitigating disease spread and reducing mortality rates \cite{chen2021earlier}.

Medical imaging modalities such as computerized tomography (CT) and chest X-ray (CXR) are important for expedited screening of COVID-19 \cite{raoof2012interpretation}. Imaging often reveals pulmonary manifestations such as multifocal ground-glass opacities and infiltrates \cite{cleverley2020role}. In current clinical practice, radiologists are required to manually inspect delineated anatomical regions in CXRs and report both normal and abnormal findings \cite{goergen2013evidence}. Given the sheer volume of data, radiologists struggle to meet workload demands. Moreover, there is a shortage of specialized radiologists which has been exacerbated by the pandemic \cite{rimmer2017radiologist}.

To address these issues, emerging automated technologies aim to facilitate expedited diagnosis and treatment of patients \cite{pang2023survey,azad2023advances,atlam2021coronavirus,keicher2023multimodal,ciarmiello2023multivariable}. With the rapid advancement of large-scale language models (LLMs), recent research has focused extensively on electronic health records (EHRs) analysis and generation of radiology reports, yielding promising results \cite{yang2022large_gatortron,papanikolaou2020dare_gpt2}. In the domains of risk assessment and patient management, as well as the conservation of healthcare resources, survival analysis techniques prove invaluable in prognostication of diseases such as COVID-19 \cite{atlam2021coronavirus}. These techniques are crucial in assisting physicians in prognostication and preliminary risk assessment. With recent advancements in deep learning and computer vision, multimodal survival analysis methods incorporating imaging, textual reports, and clinical diagnoses have the potential to offer tools for efficient physician screening  \cite{keicher2023multimodal,ciarmiello2023multivariable}. Despite this progress, critical challenges remain before clinical implementation, as described below.

\begin{figure}[!t]
\centering
\includegraphics[scale=0.137]{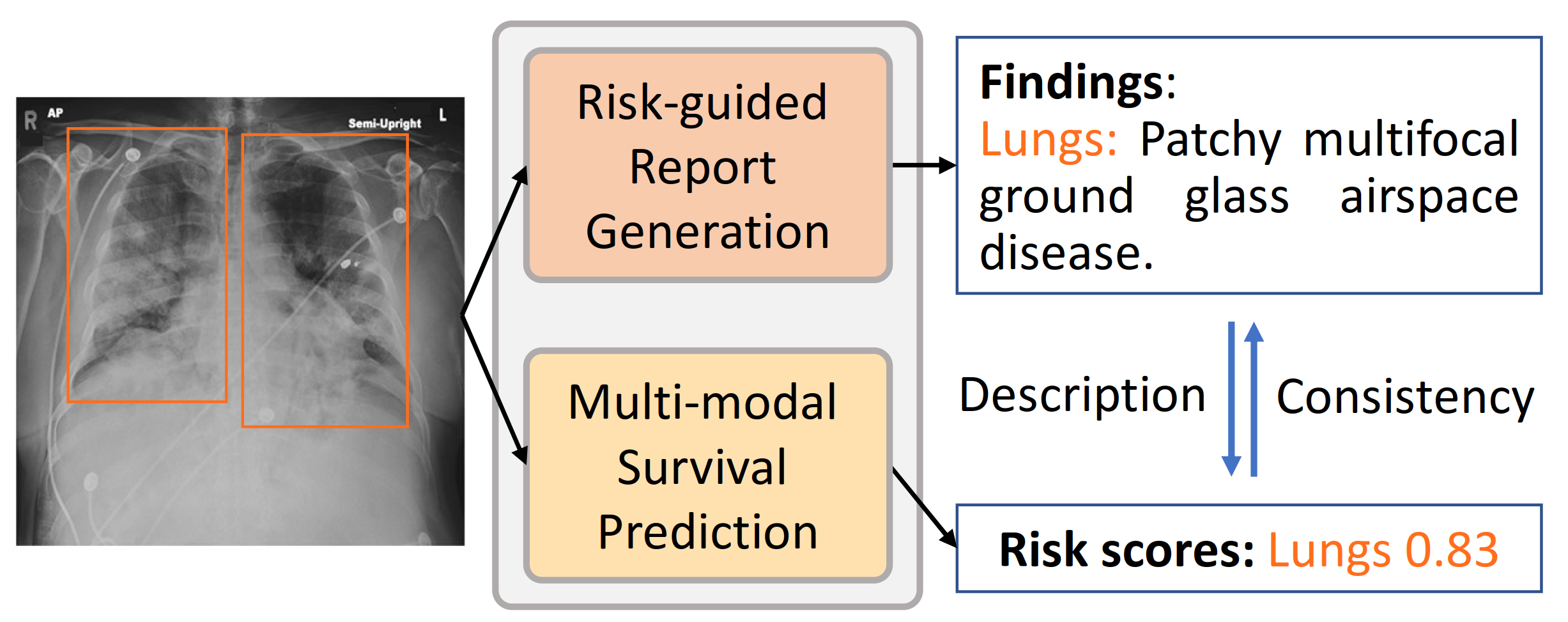}
\caption{Overview of Proposed Method: Anatomical regions detection serves as the foundation for our approach. Focusing on the detected lung region groups (represented in orange), imaging features are selected and aggregated for COVID-19 report generation and survival analysis. The generated sentence and risk score are explicitly grounded within anatomical region groups, providing mutual benefits through risk description and sentence-wise survival consistency.} \label{overview}
\end{figure}

1) Imbalanced data distribution: In COVID-19 screening, the proportion of affected to healthy individuals is low, causing distribution imbalance. This hinders accurate model attention and feature capture \cite{zhang2023intra}.

2) Visual attention deviation: Radiologists often provide general interpretations for images, describing all included items but frequently failing to highlight abnormal anatomical regions \cite{shin2016learning}. Current report generation methods rely on global visual features, leading to insufficient structured descriptions of body parts. On the other hand, current survival analyses typically offer a global perspective across modalities, but they lack the ability to link risks to specific anatomical regions.

3) Textual attention deviation: Radiology reports often contain a section describing medical observations, and similar sentences are used more frequently to describe normal areas compared to abnormal ones. This results in an overrepresentation of normal region descriptions in the report \cite{xue2018multimodal,yuan2019automatic}. The use of such imbalanced text distribution datasets for training results in the predominant generation of normal sentences, rendering the model incapable of describing specific critical anomalies.

4) Lack of model transparency and interpretability: Deep learning is advantageous due to its end-to-end processing advantages; however, the non-interpretable nature of networks and non-standardized evaluations make deep learning akin to a “black box” \cite{geis2019ethics,guidotti2018survey}. Survival analysis assesses critical risk factors, yet challenges persist in its interpretability. Ensuring transparent results and reports is crucial to allow physicians to accurately make decisions.

To ensure the model focuses adequately and accurately captures these rare yet crucial abnormal regions, this study proposes a risk-guided radiology report generation framework that conducts localized region learning of report text correlations from a risk perspective. The proposed Multi-modality Regional Alignment Network (MRANet) visually locates diagnostic areas, provides localized diagnostic descriptions, and predicts risk scores within the identified areas. Our proposed method emulates the workflow of radiologists, naturally integrating localization, description, and risk analysis \cite{pahadia2020radiology}. In MRANet, the three aspects, spanning image, text, and risk dimensions, mutually corroborate, significantly enhancing interpretability. As the overview illustrated in Fig. \ref{overview}, on the predicted regions, the predicted regional sentences not only serve as local descriptions of the image but also offer readable explanations for the risk outcomes of the multi-modality survival prediction model. Our contributions are as follows: 
\begin{itemize}
\item We present MRANet for radiology report text generation, which is capable of mutually explaining localization, text, and risk predictions. Through visual grounding and risk attention within anatomical regions, these three aspects mutually corroborate each other in generated reports and risk predictions, enhancing the model's interpretability and the readability of predictions.

\item The anatomical region detection and Multi-scale Region-feature Encoder (MRE), provides visual localization capability and aggregate the region features. The Survival-guided Sentence-feature Encoder (SSE) utilizes a risk attention mechanism, effectively and accurately capturing rare and diverse sentence feature on abnormal areas.

\item By leveraging transfer knowledge alleviates the imbalance issues of local COVID-19 data, aligning text from large-scale datasets and image models, the pretrained LLMs offering professional guidance for image feature extraction and medical report generation. This multi-stage training strategy resolves training conflicts between survival prediction and language models. 

\item We validate our model with experiments conducted on multi-center datasets. Experimental results demonstrate that our model can generate text reports related to risk, specific to localized regions, and interpret survival outcomes using the predicted sentences.
\end{itemize}

\section{Related work}
\label{sec:relatedwork}
\subsection{Survival analysis}
Survival analysis is a widely applied statistical method for assessing the time until a specific event occurs, with applications spanning various fields. Traditional approaches, such as Random Survival Forest (RSF) \cite{rsf} and the Cox Proportional Hazards (CPH) model \cite{coxmodel}, are commonly used for estimating survival risks. DeepSurv \cite{deepsurv}, an extension of the CPH model, employs a deep learning framework with the Cox negative logarithm partial likelihood loss (CoxPHloss) to capture intricate, nonlinear relationships between survival risks and covariates \cite{deepsurv}. Recent advancements emphasize the synergistic use of multiple modalities, including clinical, imaging, genetic, and time series data, to enhance disease survival analysis. Multimodal survival learning proves more robust than unimodal approaches, benefiting from the extraction of diverse information and cross-modal survival predictions \cite{keicher2023multimodal,ciarmiello2023multivariable}. 

\begin{figure*}[!t]
\centering
\includegraphics[scale=0.55]{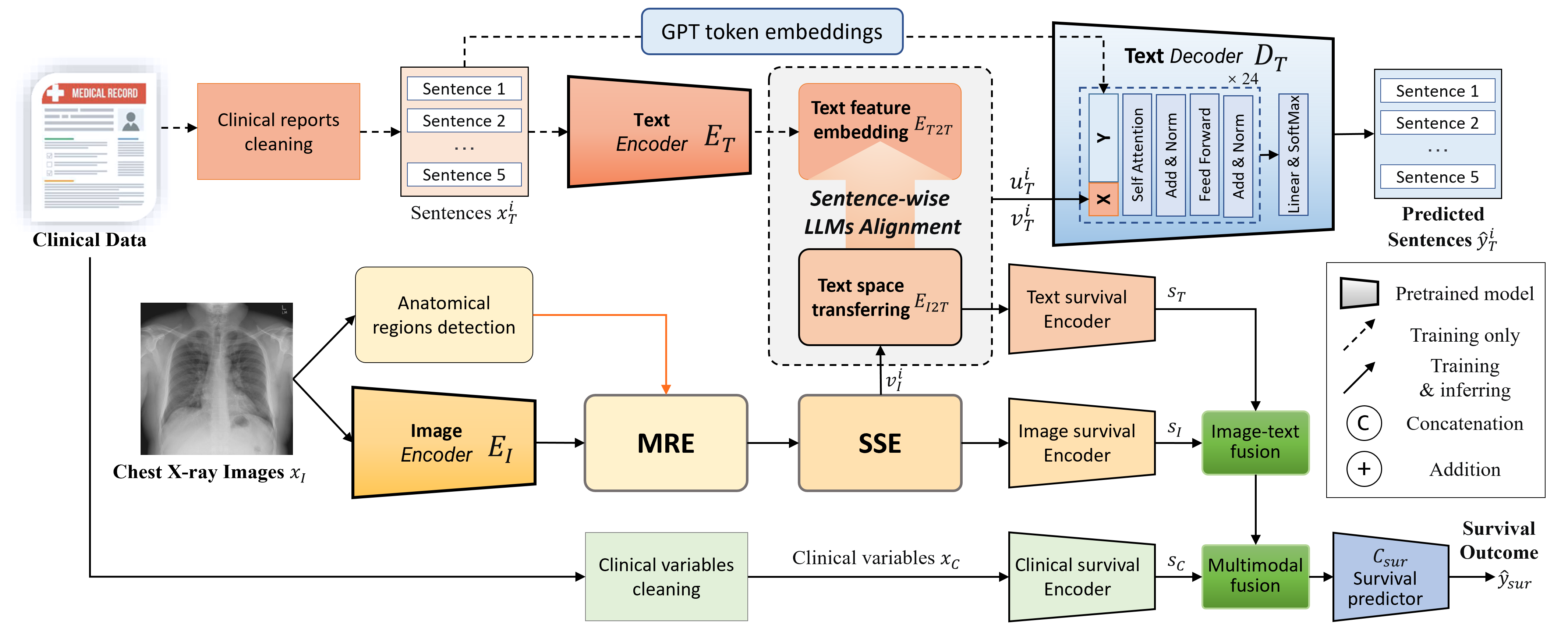}
\caption{The overview of the proposed Multi-modality Region Alignment Network (MRANet) comprising of region-based report generation with survival attention. Our framework extracts reports and clinical variables from medical data along with corresponding images for each modality—red, green, and yellow branches represent these data flows. The Multi-scale Region-Feature Encoder (MRE) aggregates local features in predicted anatomical regions to create a comprehensive feature representation for each sentence. Text decoder and encoder are LLMs used to constraint the image-to-text feature space alignment for report generation, which is embedded by Survival-guided Sentence-feature Encoder (SSE). The multi-modality survival prediction predicts the risk prediction and provides the survival-attention to the sentence feature learning in anatomical regions.} 
\label{framework}
\end{figure*}
\subsection{Radiology report generation}
In the realm of radiology report generation, the task is considerably more challenging compared to natural image captioning \cite{pang2023survey}. Despite focusing on comprehension and generating descriptive sentences, image captioning struggles to fully convey rich feature information in medical images. Radiology report generation extends beyond a single sentence, involving crafting a comprehensive paragraph with structured sentences dedicated to specific medical observations in distinct regions of radiological images \cite{filev2022structured}.

The existing approaches to medical report generation predominantly adopt encoder-decoder frameworks, evolving from LSTM-based CNN-RNN architectures to the more recent Transformer structures, benefitting from substantial datasets \cite{pang2023survey,nicolson2023improving_cvt2}. Despite the Transformer's effectiveness in processing long sequences and contextual information from extensive data, its performance is suboptimal in the context of automatically generating CXR reports due to data imbalances within the medical domain \cite{azad2023advances,seyyed2021underdiagnosis}. To address this limitation, various enhancements have been proposed, such as the incorporation of relation memory units or cross-modal memory matrices into the standard Transformer architecture \cite{chen2020generating_r2gen,chen2022cross_cmn}. Many methods introduce additional medical information to better capture anomalous text through auxiliary task learning, e.g., predicting disease keywords or tags \cite{jing2018automatic}, incorporating a topic matching mechanism \cite{yin2019automatic}, aligning disease tags with visual features \cite{you2021aligntransformer}, and introducing an additional medical knowledge graph \cite{liu2021exploring_PPKED}. In contrast, region-based generation presents an alternative approach to enhance region-focused attention through bounding box annotations. Wang et al. \cite{wang2021self} introduced a region-level extractor using a selective search algorithm, departing from global features. The RGRG \cite{tanida2023interactive_rgrg} method leverages Faster R-CNN \cite{ren2015faster} and a learnable selector for salient anatomical region object detection, enabling the independent generation of sentences that visually anchor to anatomical structures. Inspired by the working patterns of radiologists, our work aims to overcome these challenges with the multi-modal knowledge of large-scaled models. We evaluate the external survival risk to mitigate textual deviation as part of the attention mechanism for the anatomical regions. This region-aware method emphasizes learning text descriptions for specific diseases under high-risk conditions, ultimately generating well-structured radiology reports.

\section{METHODS}

\begin{figure*}[!t]
\centering
\includegraphics[scale=0.55]{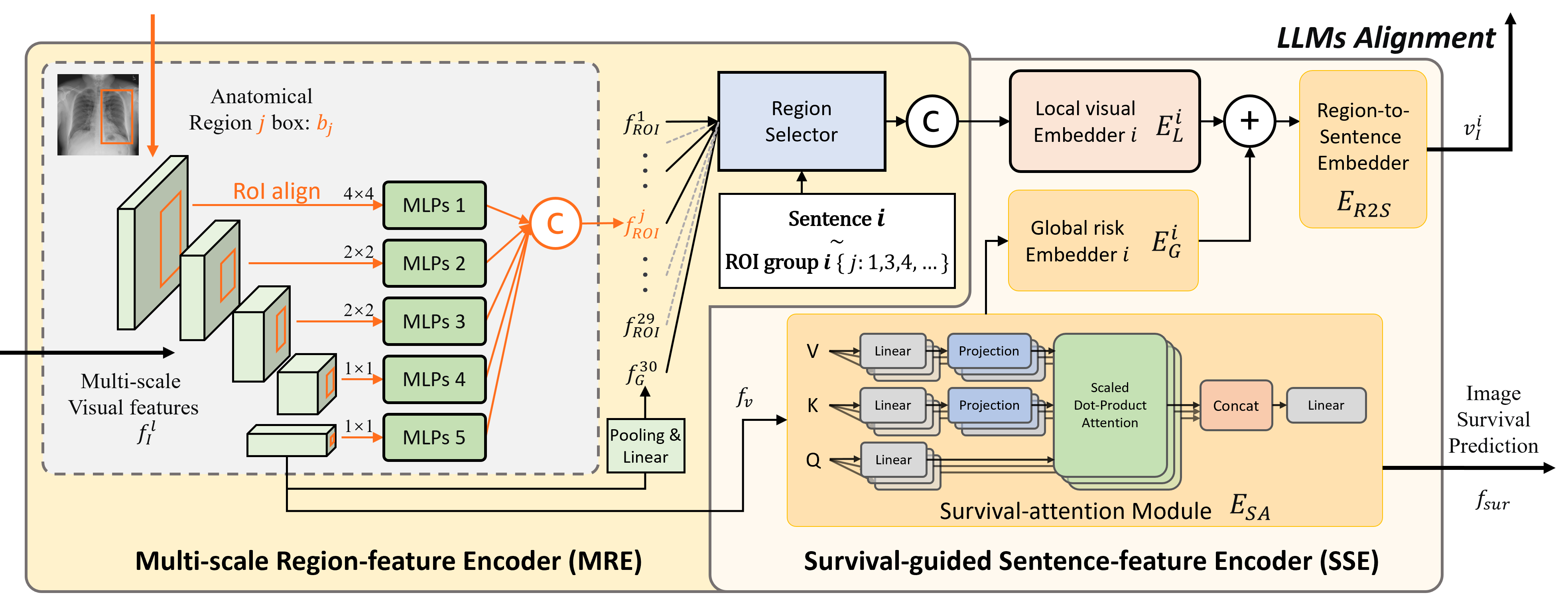}
\caption{Illustration of our proposed Multi-scale region-feature encoder (MRE) and survival-guided Sentence-feature encoder (SSE). The MRE extracts regional features from the multi-scale ResNet backbone, and aggregates by concatenating the grouped features each report sentence. The SSE takes the last visual features through a survival attention module to learn global risks. It utilizes local and global embedders to encode the visual attributes and risk attention contained in the sentences. } 
\label{MRE_SSE}
\end{figure*}

\subsection{Framework Overview}
For report generation, our model closely follows the typical workflow of a radiologist. Given a CXR image $x_I$, first, a radiologist identifies the anomalies with in anatomical areas. For each sentence, radiologists pay more attention on region with the high risk evidence and carefully summary with expertise on the anatomical regions. Finally, the structured reports includes sentences of Finding in the 4 critical regions and a global summary as Impression, represented as $x_T^i \, (i \in [1,...,5])$. \par 

The overall framework is shown in Fig. \ref{framework}. First, an object detector identifies 29 distinct anatomical regions on $x_I$. The Multi-scale Region-feature Encoder (MRE) extracts and aggregates the regions belonging to predefined sentences on the report. The Survival-guided Sentence-feature Encoder (SSE) embeds the imaging survival-attention for the regional sentences, to capture the risk-related textual expression through image-to-text transferring. Employing LLMs-Alignment, an asymmetric encoder-decoder structure constrains for text-to-text and image-to-text alignment in the shared text space. The embedded sentence features are decoded with a transformer-based language generation model. Finally, a two-stage multimodal risk feature integration is employed, leveraging three modalities additionally including clinical variables $x_C$, to predict survival outcomes through a survival predictor.

\subsection{Anatomical region detection and completion} 
For the anatomical region detector, we use Faster R-CNN \cite{ren2015faster} trained on Chest ImaGenome dataset \cite{wu2021chest}, which predicts 29 unique anatomical regions for each case. Faster R-CNN consists of a region proposal network (RPN), which generates bounding boxes of potential anatomical regions on the image feature maps. A region of interest (ROI) pooling layer maps each object proposal onto the feature maps, extracting small feature maps of uniform spatial extent. These ROI feature maps are each classified into one of the 29 anatomical region classes (and the background class) following standard procedure in Faster R-CNN. The object proposal is well-detected which should has highest probability score among the 29 class, as well as the top-1 scores for all proposals. Otherwise, the region class does not achieve the highest score for at least one proposal, it is considered as undetected.

\begin{figure}[!t]
\centering
\includegraphics[scale=0.55]{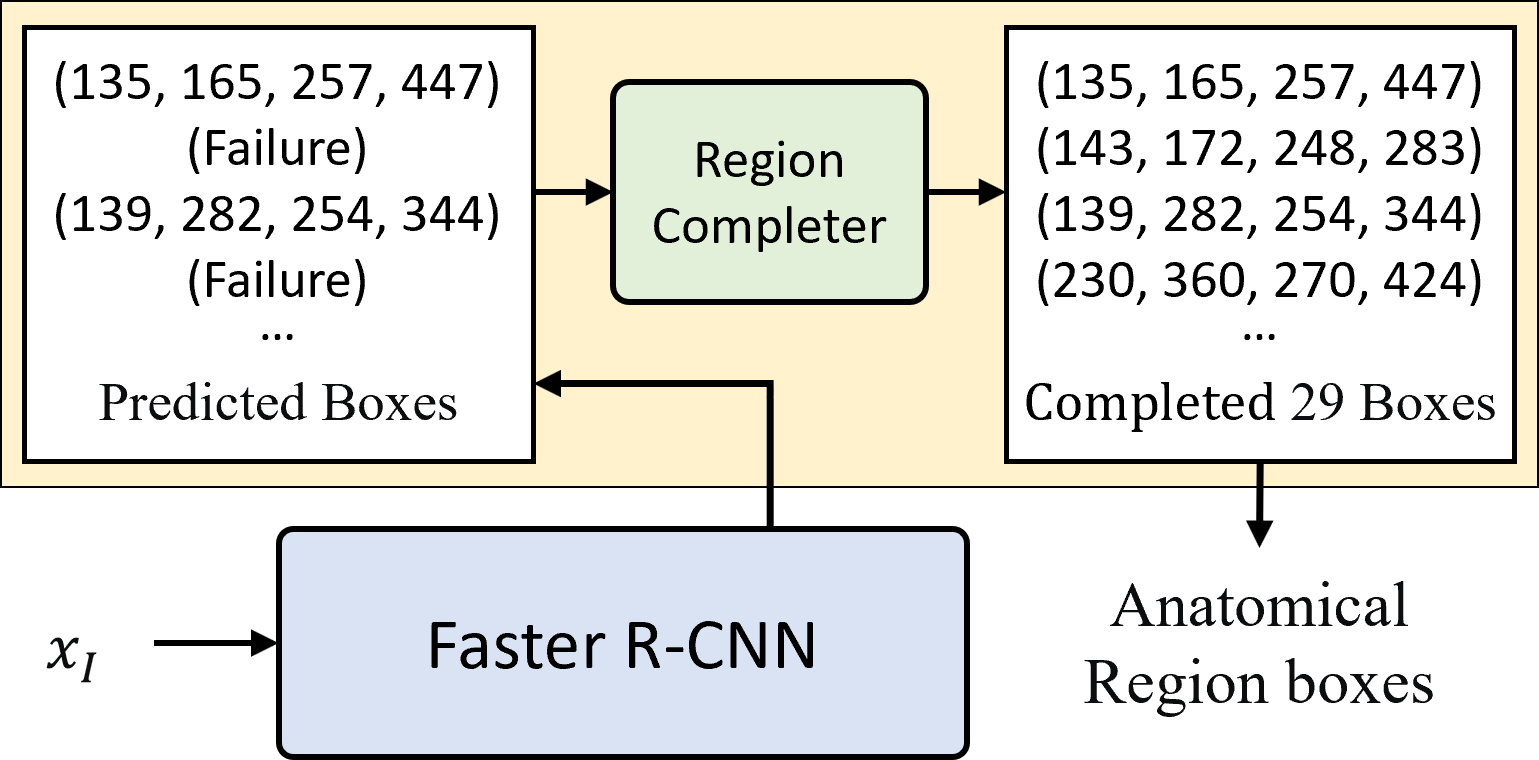}
\caption{The illustration of the anatomical region detector with Faster R-CNN and the proposed Region Completer, which corrects bounding box of the undetected region with the learned spatial coordinate pattern.} \label{completer}
\end{figure}

A compact and specialized network as shown in Fig. \ref{completer}, Region Completer is designed to rectify bounding boxes for undetected regions. The missing bounding boxes can be estimated by leveraging the initial predictions made by the region detector. The Completer is a regression network comprising a three-hidden-layer multilayer perceptron (MLPs), which is trained in the way of "Masked Autoencoders" \cite{he2022masked}. Specifically, it is trained by predicting the coordinates of several missing regions which are artificially randomly masked, thereby learning the spatial distribution of the 29 region coordinates. During the inference phase of region detection, the previously predicted coordinates utilized as fixed coordinates, which may contain undetected regions. The relative distribution correlations between known and undetected regions are inferred to predict the missing bounding boxes, thereby completing the bounding boxes for all 29 regions $b_j \, (j \in [1,...,29])$.

\subsection{Multi-scale Region-feature Encoding}
The image is first passed through the image encoding model to produce the visual features. The image encoder is ResNet-50 model pre-trained with conditional reconstruction task for both vision and language representation \cite{cheng2023prior}. Given an image $x_I$, the multi-scale visual features are extract from the 5 ResLayers of image backbone and denoted as $f_I^l = E_I(x_I)$, $\in R^{{C_l} \times {H_l} \times {W_l}}$, $l$ represents the feature scale $l \in [1,...,5]$. 

To obtain the region visual features for each sentence of radiology report, MRE module extracts and aggregates the features across multiple scales on the detected regions. Within region $b_j$, ROI align operation of $l$ scale maps local features to the corresponding ROI resolution $\in R^{{C_l} \times {H_r} \times {W_r}}$. After flattening into a one-dimensional vector, the linear transformation layer "$\textrm{MLPs} \; l$" is applied to $l$-scale feature to reduced the dimension from ${C_l} \times {H_r} \times {W_r}$ of flatten vector to 256. Then, the 256-dimensional local features of 5 scales are subsequently concatenated together into $f_{ROI}^j \in R^{1280}$. These 29 extracted ROI features represent the visual characteristics of the 29 anatomical regions. Considering the global expressions present in radiology reports, the feature $f_I^5$ from the last ResLayer is utilized to extract the global visual feature $f_{G}$ thought average pooling and a fully connected layer. This global representation is then incorporated into the set of region features, $f_{ROI}^{30} = f_{G}$, therefore $j$ belongs to the range $[1,...,29,30]$.

Normally, the sentence in a medical report identifies an observation of the anatomical regions. The MRE selects and aggregates the 30 visual features corresponding to proposal anatomical regions that may be mentioned in sentence, enabling the model to provide spatial discriminability in sentence descriptions. For a given sentence $i$, that may describe within one or more regions. These regions are categorized into the group $i$, for example, $\textbf{ROI group} \, 1 = \left \{ j: 1,3,4,... \right \}$. The Region Selector extracts the features belonging to the region group $i$ from the set of region features$f_{ROI}^j$ and performs a concatenation operation to obtain sentence-wise visual feature:
\begin{equation}
f_{ROI}^j = \displaystyle{\concat_{l=1}^{5} \textrm{MLPs}\, l \left (\textrm{ROIAlign}\, l (f_I^l)\right )}
\end{equation}
\begin{equation}
f_{MRE}^i = \displaystyle{\concat_{j \in \textbf{ROI group} \, i} f_{ROI}^j}
\end{equation}
where $ \concat$ stands for concatenation operation. $f_{MRE}$ represents the concatenated visual feature for the sentences.

\subsection{Survival-guided Sentence-feature Encoding}
SSE module is responsible for embedding the survival feature from image into sentence features. Since the deviation of descriptions for normal and abnormal findings in medical reports, SSE employs survival prediction to enhance the language generation model's risk sensitivity, guiding model to capture specific and abnormal statement and enhancing predictive accuracy for rare and high-risk case.

In SSE, the Survival-attention Module $E_{SA}$ is used to extract the risk features from the image branch, which is implemented upon the Multi-Head Attention (MHA) \cite{vaswani2017attention}. The MHA consists of $n$ parallel heads and each head is defined as a scaled dot-product attention, which is formulated as:

\begin{equation}
Att_h(X, Y) = \textrm{softmax} \left(\frac{ \left(  \left(YW^Q_h\right ) \left(XW^K_h\right ) ^T\right )}{\sqrt{d_k}}  \right )  \left(XW^V_h\right ) 
\end{equation}

\begin{equation}\label{MHA}
MHA(X, Y) = \left [\displaystyle{\concat_{h=1}^{n} Att_h(X, Y) }\right]W^O
\end{equation}
where $X$ and $Y$ denote the input matrix for Query and Key/Value. $W^Q_h$, $W^K_h$, $W^V_h$ are their learnable parameters in $h$-th head $Att_h$. $W^O$ is the weight of output linear layer. $d_k = d/n$ is the scaling factors and $d$ is dimension of embedded feature. 

We utilize Eq. \ref{MHA} to construct a self-attention pooling structure, where $E_{SA}$ extracts survival features using a self-attention encoder on the feature of last ResLayer $f_I^5 \in R^{2048 \times 7 \times 7}$. After spatial-flattening and transposition, visual sequence features $f_v \in R^{49 \times 2048}$ and their channel-wise mean vectors ${f_v}' \in R^{1 \times 2048}$, augmented with learnable positional embedding, are input into the MHA to extract survival attributes within the image features. 

\begin{equation}f_{sur} = E_{SA} \left(f_v \right ) = MHA \left( Z, Z'\right)  \end{equation}
\begin{equation}Z = f_v + Pos; \; \;Z' = {f_v}' + {Pos}'\end{equation}
where $Pos \in \mathbb{R}^{49 \times 2048}$  and $Pos' \in \mathbb{R}^{1 \times 2048}$ are the learnable spatial positional embeddings. The acquired the survival feature vectors $f_{sur} \in \mathbb{R}^{1 \times 2048}$ are utilized for survival prediction, to provide risk attention for sentence-wise feature embedding. For $i$-th sentence, the local visual embedder $E^i_L$ aggregates the sentence's visual features $f_{MRE}^i$. The global risk embedder $E^i_G$ extracts from $f_{sur}$ to associate with the context of sentence $i$ in the global view of survival prediction. The region-to-sentence embedder $E_{R2S}$ combines the survival-guided features, to obtain visual representation $\mathit{v}^i_I$. 

\begin{equation}
\mathit{v}^i_I = E_{R2S} \left ( E^i_L \left ( f_{MRE}^i  \right) + E^i_G \left ( \mathit{f}_{sur}  \right) \right) 
\end{equation}

In this way, the SSE can not only obtain sentence feature on the anatomical regions, but also align the risk feature indicating abnormal regions with the relevant descriptions, which imitates the working patterns of radiologists to assign the risk of disease to the abnormal regions when examining the abnormal regions.

\subsection{Sentence Generation}
We use the GPT-2 model \cite{papanikolaou2020dare_gpt2} as language decoder $D_T$, which is an auto-regressive neural network based on self-attention. To inject the visual features with token embeddings in the MHA of GPT-2, we follow \cite{alfarghaly2021automated} in using Pseudo Self-Attention (PSA) \cite{ziegler2019encoder}, which be formulated as:
\begin{equation}
Att_h^{PS}(X, Y) = \textrm{softmax} \left(\frac{\left(YW^Q_h\right ) \left [ \frac{XU^K_h}{YW^K_h}\right ] ^T }{\sqrt{d_k}}  \right )  \left[\frac{XU^V_h}{YW^V_h}\right ]
\end{equation}

\begin{equation}
PSA(X, Y) = \left [ \displaystyle{\concat_{h=1}^{n} Att_h^{PS}(X, Y) } \right ] W^O
\end{equation}
where $X$ represents the transferred sentence-wise visual features and $Y$ represents the token embeddings. The new parameters $U^K_h$ and $U^V_h$ are the corresponding key and value projection parameters of self attention-visual projecting. Text Decoder $D_T$ which is a GPT decoder structure, comprises a total of 24 cascaded Transformer modules including PSA layers, feedforward, residual and normalization layers. In each module, the tokens in a sequence are conditioned on previous tokens and encoded with shared visual feature $\mathit{v}^i_T$. The predicted logits are obtained in a linear layer followed by a softmax activation. Sentence predictions $\hat{y}_T^i$ are generated in an iterative process of selecting the most likely next word from the vocabulary, which based on continuously preceding tokens in the previous predicted sequence. The visual features are transferred with an embedder $E_{I2T}$ from image to text space for report generation:
\begin{equation}
\mathit{v}^i_T =  E_{I2T} \left (\mathit{v}^i_I \right); \; \; \hat{y}_T^i = D_T(\mathit{v}^i_T) 
\end{equation}

The report generation is optimized by minimizing the sum of the sentence-wise cross-entropy loss:
\begin{equation}
L_{CE}(X_T, \hat{Y}_T) = -\sum_{i=1}^{5} {\sum_{k=1}^{l_i} {x_T^{ik}log(\mathit{\hat{y}}_T^{ik})}}
\end{equation}
where $l_i$ denotes the length of words in $x_T^{i}$ of the $i$-th sentence. The cross-entropy loss for each sentence is calculated word by word until reaching the end of $x_T^{i}$.

\subsection{Image-to-text LLMs-Alignment}
To enhance the generative capacity of our model, we introduce additional knowledge as semantic constraints, incorporating a self-supervised learning paradigm based on text encoding and decoding during training. In this LLMs-Alignment procedure, GatorTron \cite{yang2022large_gatortron}, a BERT (Bidirectional Encoder Representations from Transformers) model, is employed as the Text Encoder $E_T$ from another perspective. Renowned for its exceptional capabilities in capturing intricate semantic nuances within biomedical text, the encoder can be regarded as an experienced radiologist, encapsulating the detailed information present in medical reports, and supervising for Text Decoder in the synthesis of report features into coherent textual narratives. The $E_{T2T}$ along with the $E_{i2T}$ embeds sentence features to the shared textual space for report decoding. The text encoder-decoder and the overall language generation loss are formulated as follows:

\begin{equation}
 \mathit{u}^i_T = E_{T2T} \left ( E_T \left (x_T^i \right) \right); \; \; \tilde{y}_T^i = D_T(\mathit{u}^i_T) 
\end{equation}

\begin{equation}
L_{LLMA} = L_{CE}(X_T, \hat{Y}_T) + L_{CE}(X_T, \tilde{Y}_T)
\end{equation}

During training, the reports are attractively generated on textual features $\mathit{u}^i_T$ and visual features $\mathit{v}^i_T$, with both branches sharing the Text decoder. In their shared textual space upon the GatorTron's expertise, the LLMs-Alignment guides image-to-text tranferring with clinical information perception. The alignment refines the visual generated sentences with authentic textual content within the biomedical and health domain.

\subsection{Multi-modality Survival Prediction}
The MRANet model uses the multi-modal features for survival prediction. The concatenated image-to-text features $\concat_{i=1}^{5} \mathit{v}^i_T$, $f_{sur}$ from the survival attention module, and the clinical variable $x_c$ are separately encoded through text, image, and clinical survival encoders, which are MLPs with two hidden layers, mapping multi-modal features to 2048-dimensional survival features $s_{T}$, $s_{I}$, $s_{C}$. Subsequently, a fusion module consisting of a concatenation and linear layer is applied to integrate the features of image $s_{T}$ and text $s_{I}$, followed by the 2nd fusion module to further integrate with clinical feature $s_{C}$. This two-stage fused features are then utilized to predict survival outcomes $\hat{y}_{sur}$ through a survival predictor $C_{sur}$. 

The survival prediction model is trained to differentiate between patients with critical disease and those without using CoxPHloss \cite{deepsurv}, which can handle right-censored survival data and is formulated as:
\begin{equation}
 L_{\mathrm{CoxPH}}= -\frac{1}{N_{Y_{e}=1}} \sum_{i: y_{e}^{i}=1} \left(\hat{y}^{i}_{sur} - \log \sum_{j:y_{t}^{j}>y_{t}^{i}}e^{\hat{y}^{j}_{sur}}\right) 
\end{equation}
where $\hat{y}^{i}_{sur}$ is the predicted risk of disease progression for the $i$-th patient, $y^{i}_{t}$ and $y^{i}_{e}$ are survival labels donated as the survival time and the censorship flag, respectively. $y^{i}_{t}$ is defined as the days of diagnosis to death (for $y^{i}_{e}=1$) or patient censored (for $y^{i}_{e}=0$), $N_{Y_{e}=1}$ is the number of patients with an observable event.

\section{EXPERIMENT}
\subsection{Datasets}
The Chest ImaGenome dataset \cite{wu2021chest} contains automatically constructed scene graphs for the MIMIC-CXR \cite{johnson2019mimic} dataset, consisting the images with corresponding free-text radiology reports. Each scene graph describes one frontal CXR and contains bounding box coordinates for 29 unique anatomical regions, as well as sentences describing each region if they exist in the corresponding radiology report. The clinical variables includes age, sex, temperature, oxygen saturation on room air, white blood cell count, lymphocyte count, creatinine, C-reactive protein, and comorbidities (cardiovascular disease, hypertension, chronic obstructive pulmonary disease, diabetes, chronic liver disease, chronic kidney disease, cancer, and HIV status) \cite{jiao2021prognostication}. We use the split provided by the dataset: 166,512 training, 23,952 validation, and 47,389 test images. 

\begin{figure*}[!t]
\centering
\includegraphics[scale=0.55]{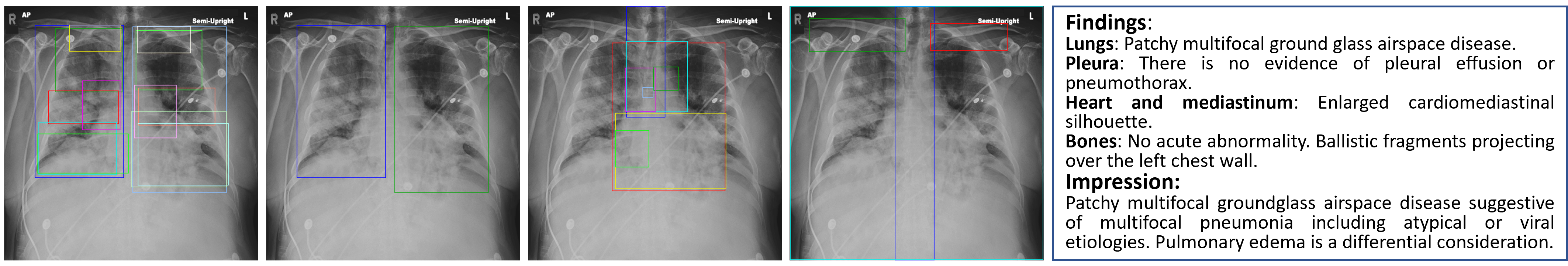}
\caption{An example of Chest X-Ray image with detected anatomical regions and the corresponding structural report. The 4 sentences (Lungs, Pleura, Heart and mediastinum, and Bones) of \textbf{Findings} are the summaries on the regions in 4 groups as illustrated on left 4 figures. The sentence of \textbf{Impression} is the description on whole image.} \label{sample}
\end{figure*}

A retrospective review was conducted to identify COVID-19 patients with imaging between March 2020 and July 2020 at Brown Affiliated Hospitals (Brown) and the Hospital of the University of Pennsylvania (Penn). The Brown-COVID dataset contains 1021 front-view CXR and corresponding reports. Different from the free-text reports from MIMIC-CXR dataset, our structured reports include 5 sentences as shown in Fig. \ref{sample}; each are well defined for 4 Findings which included details for their defined regions and the Impression, which serves as a summary of the report. The concatenation was used as the target report in experiments. The survival labels include mortality event and the length of time from COVID-19 test to date of follow-up or mortality. The Penn-COVID dataset contains 2879 images and survival labels, but lacks radiology reports. We conducted our model evaluations in a multi-center setting to ensure robustness and generalizability. The Brown-COVID dataset was divided into three sets with 70\% for training, 10\% for validation, and 20\% for internal testing. The Penn-COVID dataset was used external validation of survival prediction. The study and collection of dataset have been approved by Lifespan Institutional Review Board, Providence, RI, USA (IRB board protocol: [1584610-18]; approval date: 10/09/2023). 

\subsection{Training and Inference}
The task-specific training protocol is designed to cater to the unique target of individual modules.
\subsubsection{Training of anatomical region detection} We use the images and 29 anatomical region boxes of Chest ImaGenome dataset to train the object detector and region completer, as illustrated in Fig. \ref{completer}. The object detector loss $L_{obj}$ of Faster R-CNN encompasses both proposal classification loss and boundary box regression loss. Additionally, the region completer loss calculates the Mean Squared Error (MSE) to evaluate the predictive completion from the artificially masked input coordinates.

\subsubsection{Training of report generation and survival prediction} 
This training process is designed into three distinct steps, reducing computation conflicts between the training tasks. The training of survival prediction involves capturing inductive statistical patterns, requiring the utilization of a larger batch size for more accurate gradient descent optimization. However, the report generation model has a large number of parameters, allowing only a small batch size to match the computing resources. The three-stage task orchestration follows the optimization sequence of:
\begin{renum}
\item Survival attention module. We train the survival attention module to predict the image-based survival outcome $\hat{y}_{sur\textrm{-}I} = C{sur}(E_{SA}(E_I(x_I)))$ and optimize with $L_{\mathrm{CoxPH}}$. This enables the attention module to extract survival-related features from image features. 

\item Survival-guided report generation. Reload and fix the parameters of $E_{SA}$ to extract risk attention for the image-to-text feature encoding. $L_{LLMA}$ serves as the objective function, optimizing the encoding of region visual features in SSE and MRE, along with the weights for text space mappings $E_{T2T}$, $E_{I2T}$. 

\item Multi-modal survival prediction. The fused survival outcomes based on two-stage fusing are optimized with target function $L_{\mathrm{CoxPH}}$. Only the survival encoders and fusion modules of the three modalities are updated during this process. 
\end{renum}

\subsubsection{Test setting}
While inference, MRANet model detects the 29 anatomical region and outputs the structured reports include 5 independently generated sentences associate with regional survival risk attention, and global multi-modal risk scores.

\subsection{Experimental Settings}
\subsubsection{Evaluation Metrics} In this paper, natural language generation (NLG) metrics are used to measure the quality of the generated reports: BLEU \cite{papineni2002bleu}, METEOR \cite{banerjee2005meteor}, ROUGE-L \cite{chin2004rouge} and and CIDEr-D \cite{vedantam2015cider}. To be specific, BLEU-n (Bi-Lingual Evaluation Understudy) is used to evaluate the n-gram overlap between the generated and ground-truth reports. In contrast, WordNet is used in METEOR (Metric for Evaluation of Translation with Explicit ORdering) to retrieve all possible synonyms of the extracted concepts between uni-grams. Besides, ROUGE (Recall-Oriented Understudy for Gisting Evaluation) measures the longest overlapping units such as n-grams between the caption generation and the ground truth. CIDEr (Consensus-based Image Description Evaluation) emphasizes diversity and richness of expression by quantifying the consensus between generated and reference texts.

NLG metrics quantify the resemblance of generated reports to reference standards, they may not consistently reflect diagnostic precision. To address this, the Clinical Efficacy (CE) metrics \cite{liu2019clinically} make use of the CheXpert labeler \cite{irvin2019chexpert}, which is a tool developed to extract 14 observations from the reports of the CheXpert dataset (12 of which are abnormalities, normal and unmentioned). To calculate the CE metric, 14 observations are first extracted from the generated and ground-truth reports. Following this, the precision, recall, and F-1 score between the observations of the generated and ground-truth reports are calculated to give the scores of the CE metric. 

Survival analysis and time-to-event predictions commonly report results using the Concordance Index (C-index) \cite{cindex}, which assesses a model’s ability to correctly order pairs of subjects based on their predicted risk scores and corresponding event times (e.g., death, failure). 

\begin{table*}[!ht]
    \centering
    \caption{ C-index of survival prediction methods comparison. }
    \label{compare_sp}
    \renewcommand\arraystretch{0.7}
    \begin{tabular}{lcccccccc}
    \toprule[ 0.9 pt]
        Method  & Brown-img & Brown-text & Brown-clin & Brown-fuse & Penn-img & Penn-text &  Penn-clin & Penn-fuse \\ 
        \midrule [ 0.6 pt]
        RSF-CoxPH & 0.752 & 0.705 & 0.688 & 0.755 & 0.709 & - & 0.664 & - \\ 
        CPH-CoxPH & 0.688 & 0.728 & 0.729 & 0.723 & 0.582 & - & 0.680 & - \\ 
        Vanilla-CoxPH & 0.737 & 0.746 & 0.739 & 0.750 & 0.713 & - & 0.703 & - \\ 
        Multi-Fuse & - & - & - & 0.757 & - & - & - & - \\ 
        Ours & - & - & - & $\mathbf{0.813}$ & - & - & - & $\mathbf{0.739}$ \\ 
    \bottomrule[ 0.9 pt]    
    \end{tabular}
\end{table*}

\subsubsection{Backbone models}
We integrated large-scale trained models for imaging and textual learning, to incorporate well-learned knowledge from other domains. 
\begin{itemize}
\item Image encoder: PRIOR \cite{cheng2023prior}, a Medical Vision Language Pretraining model, was used to obtain the fine-grained representation on MIMIC-CXR \cite{johnson2019mimic}. ResNet50, the image backbone in PRIOR, was used for image feature encoding, which learned the visual feature from the large-scale CXR dataset. We use the checkpoint as pretrained model of the image encoder.
\item Text encoder of LLMs-Alignment: GatorTron \cite{yang2022large_gatortron} is BERT-style model and self-supervised trained on over 90 billion words medical terminologies and clinical texts in EHRs. This model extract meaningful clinical representation and demonstrated superior performance on 5 clinical NLP tasks. To extract the sentence-wise textual features from Brown-COVID reports, GatorTron-medium which includes 3.9 billion parameters, was used as the text encoder in MRANet.
\item Text decoder: For the text generation from sentence-wise visual features, we used GPT-2 with 355 million parameters that were fine-tuned for medical report generation on the 500k PubMed abstracts \cite{papanikolaou2020dare_gpt2}. This model offered the effectiveness knowledge of using curated biomedical corpora for CXR report modelling. We replaced the MHA with SPA and only trained the projection parameters for visual injecting in GPT decoder. 
\end{itemize}
\subsubsection{Implementation details}
We used the checkpoint of RGRG for region detection and trained the region completer individually with a special setting (batch size: 2000, learning rate: $10^{-3}$). The region groups in Fig. \ref{sample} show grounding of the potential combinations of regions aligned with the description in the 4 finding sentences. 

For visual feature encoding, the CXRs are first resized to $224\times 224$ and used to extract multi-scale features from ResNet50, with dimensions of 64, 256, 512, 1024, 2048. Then, the ROI align processes with kernel size of 4,2,2,1,1, embed to 256 dimensions on the regional multi-scale feature. We trained 100 and 20000 epochs for survival task and report generation on Brown-COVID dataset, with a learning rate of $10^{-4}$ and $10^{-5}$, respectively. We adopted AdamW as the optimizer with a batch size of 16. The number $n$ of attention heads per MHA and SPA layer was 8. The MRANet framework was implemented on the survival analysis packages and Huggingface on 2 NVIDIA GV100 32GB GPUs.

\subsubsection{Ablation settings of MRANet}
To respectively investigate the influence of MRE and SSE module and LLMs-Align constraint on report generation and survival prediction, the following ablation settings were used:
\begin{itemize}
\item BASE: This is the backbone of MRANet, including image encoder, text decoder and multi-modal survival prediction in our full model. The visual features for 5 sentences are embedded directly using the global feature derived from last ResLayer, the global features also serves as risk feature for image-based survival prediction tasks. 
\item BASE + MRE: This ablation incorporates the anatomical region detection and multi-scale region image encoding block into the base model. In this setting, 5 visual features are embedded and aggregated on the region feature derived from multi-scale ResLayers. 
\item BASE + SSE: To study the effect of survival attention from images to sentence embeddings, we introduce the Survival Attention module into the base model. The 5 visual features in this setting are embedded from global feature and then integrated with survival-attentional information derived from the image, to indicate the risk attribute in textual descriptions.
\item BASE + MRE + SSE: This setting combines both MRE and SSE modules during sentence feature encoding, allowing us to investigate their joint impact.
\item BASE + MRE + SSE + LLMs-Align: In this last ablation setup, we introduce the constraint for image-to-text space transferring in addition to the previous components. This enables us to analyze how the knowledge of biomedical concepts from GatorTron can enhance report generation and survival prediction performance.
\end{itemize}

\begin{table*}[!ht]
    \centering
    \caption{\label{NLG} NLG metric and CE metric scores on the Brown-COVID dataset.}
    \setlength{\tabcolsep}{4.5pt}
    \renewcommand\arraystretch{0.7}
    \begin{tabular}{lcccccccc|ccc}
    \toprule[ 0.9 pt]
        Method & Year & BLEU-1 & BLEU-2 & BLEU-3 & BLEU-4 & METEOR & ROUGE-L & CIDEr & Precision & Recall & F-1 \\ 
        \midrule [ 0.6 pt]
        R2Gen \cite{chen2020generating_r2gen} & 2020 & 0.271 & 0.116 & 0.063 & 0.036 & 0.110 & 0.186 & 0.029 & 0.359 & 0.278 & 0.294 \\ 
        CMN \cite{chen2022cross_cmn} & 2021 & 0.242 & 0.113 & 0.064 & 0.037 & 0.098 & 0.189 & 0.027 & 0.306 & 0.292 & 0.283\\ 
        CvT-212DistilGPT2 \cite{nicolson2023improving_cvt2} & 2022 & 0.210 & 0.105 & 0.063 & 0.040 & 0.093 & 0.175 & 0.059 & 0.306 & 0.292 & 0.285\\ 
        RGRG \cite{tanida2023interactive_rgrg} & 2023 & 0.309 & 0.179 & 0.116 & 0.076 & 0.117 & 0.222 & 0.092 & 0.282 & 0.342 & 0.293\\ 
        \midrule [ 0.6 pt]
        BASE & Ours & 0.475 & 0.383 & 0.327 & 0.288 & 0.241 & 0.452 & 0.396 & 0.594 & 0.546 & 0.537 \\ 
        BASE + MRE & Ours & 0.498 & 0.403 & 0.346 & 0.306 & 0.252 & 0.464 & 0.271 & 0.624 & 0.585 & 0.568\\ 
        BASE + SSE & Ours & 0.485 & 0.395 & 0.343 & 0.305 & 0.246 & 0.470 & 0.417 & 0.599 & 0.551 & 0.537\\ 
        BASE + MRE + SSE & Ours & 0.504 & 0.416 & 0.363 & 0.324 & 0.252 & $\mathbf{0.482}$ & 0.346 & 0.633 & 0.590 & 0.576\\ 
        \makecell[l]{BASE + MRE + SSE \\ \ \ \ \ \ \ \ \  + LLMs-Align} & Ours & $\mathbf{0.504}$ & $\mathbf{0.416}$ & $\mathbf{0.363}$ & $\mathbf{0.324}$ & $\mathbf{0.255}$ & 0.481 & $\mathbf{0.448}$ & $\mathbf{0.661}$ & $\mathbf{0.611}$ & $\mathbf{0.600}$ \\ 
    \bottomrule [ 0.9 pt] 
    \end{tabular}
\end{table*}

\subsection{Survival prediction performance}
We conducted experiments on survival prediction, comparing our proposed method with widely used baseline approaches. The first three methods listed in Table \ref{compare_sp} include three separate survival predictors that are trained to estimate risk scores based on image, ground-truth text report, and clinical variables, respectively. These models utilize image encoders and text encoders in MRANet for feature extraction; their respective survival predictors are RSF \cite{rsf}, CPH model \cite{coxmodel}, and Vanilla \cite{deepsurv}. Subsequently, the risk scores from each modality are combined using a half-parametric linear analysis with the CoxPH model \cite{coxmodel} to obtain the fused risk score. The Multi-Fuse model in Table \ref{compare_sp} is based on the multi-modal survival prediction structure of MRANet that utilizes one survival model for predicting the fused risk scores on the fused survival feature. Given the absence of text reports in the Penn dataset, there are missing values.

As presented in Table.\ref{compare_sp}, MRANet achieves C-index values of 0.813 and 0.739 for Brown-COVID and Penn-COVID datasets, respectively, surpassing competing methods. The models fused with CoxPH demonstrated the presence of complementary survival-related information between different modalities. Across different modalities, RSF and Cox show relatively better performance as survival predictors. Meanwhile, Vanilla outperforms traditional survival prediction models in clinical experiments (0.739 and 0.703 for Brown and Penn) while maintaining a more stable overall performance with superior generalization on external validation sets. Compared to the single-modal survival prediction then late fusion strategy employed by Vanilla-CoxPH, Multi-Fuse is more effective in analyzing multi-modal features through learnable fusion. Although our proposed MRANet does not rely on real text and employs image-to-text features for survival prediction, it surpasses the performance of existing methods. This highlights that semantic information contained within language can be harnessed to improve survival analysis in generated reports.

\begin{table}[!h]
    \centering
    \caption{ C-index of single modality backbones.}
    \label{single_sp}
    \setlength{\tabcolsep}{11 pt}
    \renewcommand\arraystretch{0.7}
    \begin{tabular}{lcccc}
    \toprule[ 0.9 pt]
        Modal & Pretrain & Fix $E_I$ & Brown & Penn \\ 
        \midrule [ 0.6 pt]
        Image & No & $\checkmark$ & 0.603 & 0.530 \\ 
        Image & ImageNet & $\times$ & 0.696 & 0.615 \\ 
        Image & ImageNet & $\checkmark$ & 0.718 & 0.619 \\ 
        Image & PRIOR \cite{cheng2023prior} & $\times$ & 0.722 & 0.689 \\ 
        Image & PRIOR \cite{cheng2023prior} & $\checkmark$ & $\mathbf{0.737}$ & $\mathbf{0.713}$ \\ 
        \midrule [ 0.6 pt]
        Text & No & $\times$ & 0.746 & - \\ 
        Clin & No & $\times$ & 0.739 & 0.703 \\ 
    \bottomrule[ 0.9 pt]    
    \end{tabular}
\end{table}

Table \ref{single_sp} shows survival prediction performance on single modalities using extracted features from images, reports, and clinical variables as inputs to a single-modal MLP model. We compare image-based performance against various ResNet50 checkpoints while conditioning frozen parameters of $E_I$. As demonstrated in the table, PRIOR from large-scale pretraining on MIMIC-CXR data—outperforms ImageNet-pretrained models for survival prediction. Furthermore, when image encoder parameters are fixed, superior performance is achieved. This highlights that, training with smaller datasets leads to overfitting of the survival model, leading to reduced generalization on two test sets. The last three rows of Table \ref{single_sp} correspond to Vanilla-CoxPH's unimodal survival prediction performance. Merely combining these modalities does improve for survival prediction.

\begin{table}[!h]
    \centering
    \caption{\label{abalation_sp} C-index of ablation studies.}
    \setlength{\tabcolsep}{11 pt}
    \renewcommand\arraystretch{0.7}
    \begin{tabular}{lcc}
    \toprule[ 0.9 pt]
        Method & Brown & Penn \\ 
        \midrule [ 0.6 pt]
        BASE  & 0.787 & 0.718 \\ 
        BASE + MRE & 0.777 & 0.724 \\ 
        BASE + SSE  & 0.803 & 0.733 \\ 
        BASE + MRE + SSE & 0.810 & 0.734 \\ 
        BASE + MRE + SSE + LLMs-Align & $\mathbf{0.813}$ & $\mathbf{0.739}$ \\ 
    \bottomrule[ 0.9 pt]    
    \end{tabular}
\end{table}

To verify the survival prediction effectiveness of each component, we perform individual experiments on our the ablated variations. Table \ref{abalation_sp} shows that the steady improvement on each component demonstrates the effectiveness, while highlighting the overall coherence of the module composition within the proposed model.

\subsection{Results on report generation}
The experimental results of the report generation on the Brown-COVID datasets are presented in Table \ref{NLG}, including a comparison with state-of-the-art (SOTA) models and ablation studies for MRANet. We used the models trained on MIMIC-CXR as baselines and directly applied them on images of Brown-COVID. In generating radiology reports, our proposed model is either superior or comparable to previous models for NLG and CE metrics. Specifically, our methods achieves the goal of structure report generation when fine-tuned locally on the small-scale dataset, surpassing all large-scaled models optimized for free-text report generation.

Regarding BLEU scores, MRANet surpasses recent state-of-the-art models \cite{tanida2023interactive_rgrg}, achieving superior performance on report summarization. Notably, RGRG leverages abnormal region detection to capture rare descriptions in reports, which enhances regional detail description abilities and increases BLEU scores compared to other baselines. Compared to the region-based baseline, MRANet's base model exhibits a remarkable 70\% improvement in BLEU-4 score. This significant difference stems from our method's utilization of specified regional cluster features, leading to higher anatomy sensitivity and enhanced structural predictability within generated reports. Our final model also excels on the METEOR metric and ROUGE scores for summarization and stemming/synonymy matching as compared to previously reported best results in medical report generation.

MRANet achieves an average 50\% increase in CE metrics compared to baseline methods. Global description approaches outperform local baselines due to RGRG's limited descriptions of only selected areas, which may not be sufficiently comprehensive. If high-risk regions go undetected or are not chosen for description, crucial image features can be overlooked, diminishing disease description performance.

Our method generates sentences based on the aggregation of multiple presumptively defining regions, incorporating all possible descriptive areas for comprehensive learning about region relationships and their importance through survival attention. This approach allows our model to focus on rare details while providing a more complete description compared to the region-based baseline. Furthermore, integrating GatorTron knowledge constraints with LLMs-Align significantly improves Precision (8.5\%), Recall (5.9\%), and F1 score (6.2\%) for medical report generation tasks, highlighting the positive impact of additional clinical knowledge in enhancing the model's ability to capture clinically relevant details within generated reports.

\subsection{Qualitative analysis}
\begin{figure}[!t]
\centering
\includegraphics[scale=0.58]{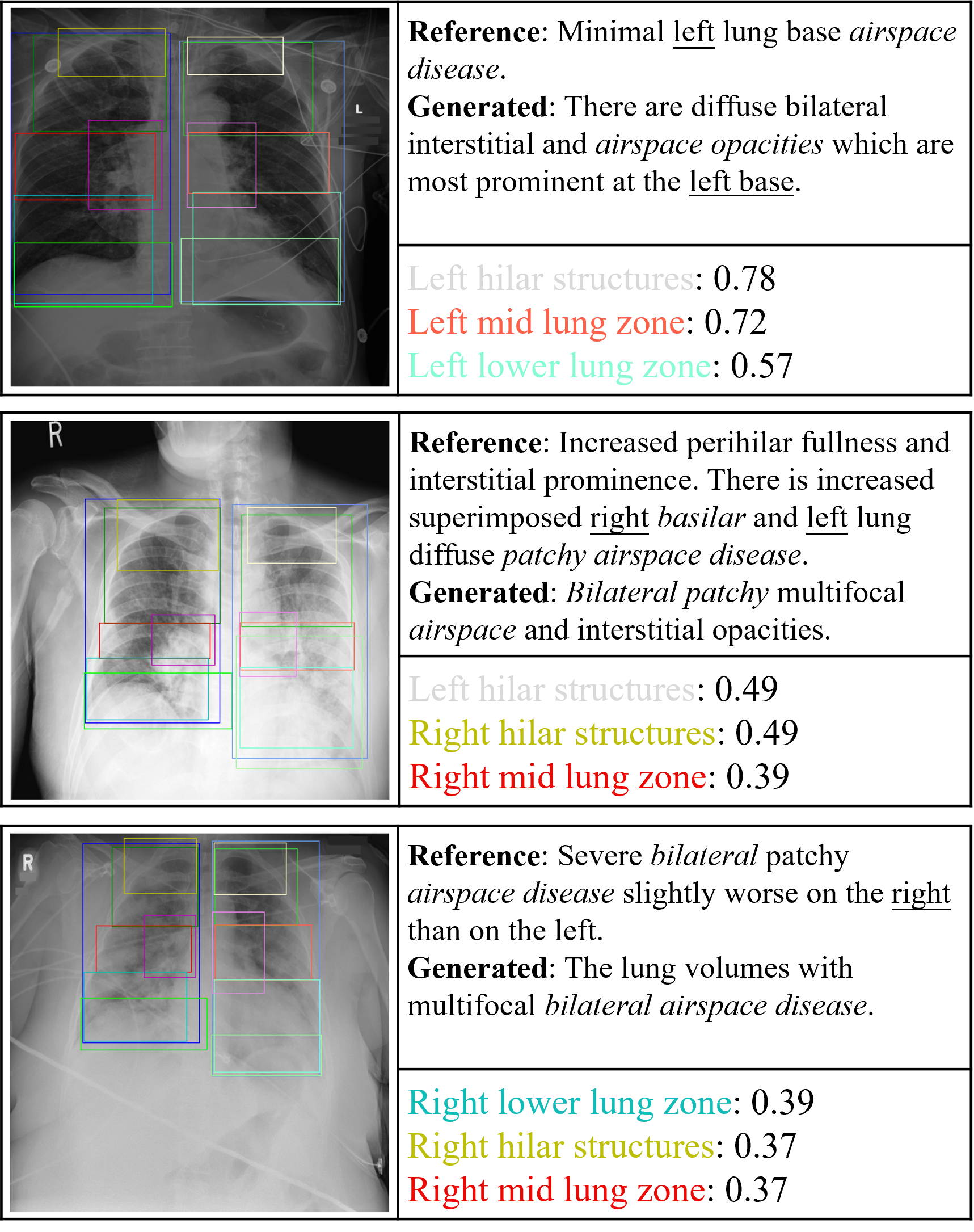}
\caption{Qualitative results of Lung-region analysis including predicted anatomical regions, the corresponding generated sentences and the regional risk attention scores. In textual reports, underlined terms pertain to localization, whereas italicized terms relate to disease description. Risk attention scores reflect the importance of high-risk regions during sentence generation, ranking the top-3 regional scores.} 
\label{vis}
\end{figure}

The proposed MRANet constitutes an integrated three-task model, harmoniously merging anatomical region detection, radiology report generation, and survival prediction. This section highlights the multi-task outcomes for lung regions and conduct a qualitative analysis to assess their correlations. As the results of Brown-COVID texting set demonstrated in Fig. \ref{vis}, the predicted bounding boxes accurately localize lung and subregions across diverse gray-level images, underscoring the object detector's robust performance. The model generates accurate descriptions on the disease-affected regions, which can correctly describe the location and diseases. The missing disease grounding descriptions can result from having limited training data, which may not provide enough location context for the language model to specific region. 

Risk attention scores are obtained utilizing Grad-CAM tools \cite{selvaraju2017grad} to evaluate each region feature's contribution towards survival outcomes within the local visual embedder encoding process. Calculating mean gradient values of predicted multimodal risk scores as weights for region features, regional attention scores result from regional-wise sum of weighted feature map and ReLU activation application. By re-weighted with global survival score, we observed that abnormal areas consistently yield higher risk attention scores compared to normal ones, offering diverse reference to text descriptions. For instance, the 3rd example despite an omission in spatial positioning within generated sentences, elevated risk attention scores can still assist in corroborating the accuracy of "worse on the right than on the left" descriptions. 

\section{Limitations}
While the proposed MRANet achieves excellent performance, there remain several limitations. Firstly, it is difficult to directly compare the report generation capability of our method to competitors’, as competitor methods produce free-text reports focusing solely on anomalous regions, while our approach incorporates descriptions for all regions, including normal ones. The discrepancy results in biased NLG evaluation metrics, which inadequately captures the model's precision in disease description. The more accurate assessment is conferred by CE metrics, substantiating our exceptional performance. 

Secondly, our datasets lack region annotations; therefore, real region boundary boxes are unavailable for direct evaluations of detectors and region completer. As such, the evaluation of the predicted bounding boxes can only be measured improvement in the ablation studies which includes MRE module. Given the limited size of our collected dataset, which may be the special cases compared to large-scale public datasets, our method is best suited for structured reports where each sentence has a specific described region.

Finally, our research concentrates on enhancing explainability within model and result prediction dimensions, providing technically justifiable insights. However, real-world clinical effectiveness has yet to be empirically demonstrated. Predictive credibility is an essential aspect of AI models employed in medical studies that warrants further investigation for ensuring trustworthiness in practical applications.

\section{Conclusion}
In this paper, we introduce a novel, explainable model for radiology report generation and survival prediction that focuses on high-risk regions. The robust anatomical regions are generated by learning spatial correlation in the detector, providing a visual grounding ability for region-specific descriptions. The visual features of each region are embedded and aggregated using a survival attention mechanism, providing spatially and risk-aware features for sentence encoding. A novel COVID-19 CXR generation style, resembling comprehensive supervision by an expert clinician, effectively constrains the image-to-text transfer process, enhancing visual feature learning while reconstructing sentences rich in clinical detail. The two tasks are complementary integrated to provide text interpretations to survival analysis, improving clinical interpretation. Moreover, consistent risk prediction between generation and analysis enhances model credibility through cross-validation. Spatially indicative language predictions and risk assessments are achieved by incorporating local features while maintaining global coherence across the two tasks. The effectiveness of this integrated system is demonstrated in multi-center experiments, validating both its overall performance and the rationality of each module’s composition within the model. Future work may expand on our findings to improve technical transparency and clinical interpretability in healthcare AI models.

\section*{REFERENCES}
\bibliographystyle{IEEEtran}
\bibliography{ref}

\begin{thebibliography}{10}
\providecommand{\url}[1]{#1}
\csname url@samestyle\endcsname
\providecommand{\newblock}{\relax}
\providecommand{\bibinfo}[2]{#2}
\providecommand{\BIBentrySTDinterwordspacing}{\spaceskip=0pt\relax}
\providecommand{\BIBentryALTinterwordstretchfactor}{4}
\providecommand{\BIBentryALTinterwordspacing}{\spaceskip=\fontdimen2\font plus
\BIBentryALTinterwordstretchfactor\fontdimen3\font minus \fontdimen4\font\relax}
\providecommand{\BIBforeignlanguage}[2]{{%
\expandafter\ifx\csname l@#1\endcsname\relax
\typeout{** WARNING: IEEEtran.bst: No hyphenation pattern has been}%
\typeout{** loaded for the language `#1'. Using the pattern for}%
\typeout{** the default language instead.}%
\else
\language=\csname l@#1\endcsname
\fi
#2}}
\providecommand{\BIBdecl}{\relax}
\BIBdecl

\bibitem{world2020announces}
I.~World Health~Organization \emph{et~al.}, ``Who announces covid-19 outbreak a pandemic,'' 2020.

\bibitem{chen2021earlier}
Y.-J. Chen, W.-H. Jian, Z.-Y. Liang, W.-J. Guan, W.-H. Liang, R.-C. Chen, C.-L. Tang, T.~Wang, H.-R. Liang, Y.-M. Li \emph{et~al.}, ``Earlier diagnosis improves covid-19 prognosis: a nationwide retrospective cohort analysis,'' \emph{Annals of Translational Medicine}, vol.~9, no.~11, 2021.

\bibitem{raoof2012interpretation}
S.~Raoof, D.~Feigin, A.~Sung, S.~Raoof, L.~Irugulpati, and E.~C. Rosenow~III, ``Interpretation of plain chest roentgenogram,'' \emph{Chest}, vol. 141, no.~2, pp. 545--558, 2012.

\bibitem{cleverley2020role}
J.~Cleverley, J.~Piper, and M.~M. Jones, ``The role of chest radiography in confirming covid-19 pneumonia,'' \emph{bmj}, vol. 370, 2020.

\bibitem{goergen2013evidence}
S.~K. Goergen, F.~J. Pool, T.~J. Turner, J.~E. Grimm, M.~N. Appleyard, C.~Crock, M.~C. Fahey, M.~F. Fay, N.~J. Ferris, S.~M. Liew \emph{et~al.}, ``Evidence-based guideline for the written radiology report: Methods, recommendations and implementation challenges,'' \emph{Journal of medical imaging and radiation oncology}, vol.~57, no.~1, pp. 1--7, 2013.

\bibitem{rimmer2017radiologist}
A.~Rimmer, ``Radiologist shortage leaves patient care at risk, warns royal college,'' \emph{BMJ: British Medical Journal (Online)}, vol. 359, 2017.

\bibitem{pang2023survey}
T.~Pang, P.~Li, and L.~Zhao, ``A survey on automatic generation of medical imaging reports based on deep learning,'' \emph{BioMedical Engineering OnLine}, vol.~22, no.~1, pp. 1--16, 2023.

\bibitem{azad2023advances}
R.~Azad, A.~Kazerouni, M.~Heidari, E.~K. Aghdam, A.~Molaei, Y.~Jia, A.~Jose, R.~Roy, and D.~Merhof, ``Advances in medical image analysis with vision transformers: A comprehensive review,'' \emph{arXiv preprint arXiv:2301.03505}, 2023.

\bibitem{atlam2021coronavirus}
M.~Atlam, H.~Torkey, N.~El-Fishawy, and H.~Salem, ``Coronavirus disease 2019 (covid-19): Survival analysis using deep learning and cox regression model,'' \emph{Pattern Analysis and Applications}, vol.~24, pp. 993--1005, 2021.

\bibitem{keicher2023multimodal}
M.~Keicher, H.~Burwinkel, D.~Bani-Harouni, M.~Paschali, T.~Czempiel, E.~Burian, M.~R. Makowski, R.~Braren, N.~Navab, and T.~Wendler, ``Multimodal graph attention network for covid-19 outcome prediction,'' \emph{Scientific Reports}, vol.~13, no.~1, p. 19539, 2023.

\bibitem{ciarmiello2023multivariable}
A.~Ciarmiello, F.~Tutino, E.~Giovannini, A.~Milano, M.~Barattini, N.~Yosifov, D.~Calvi, M.~Setti, M.~Sivori, C.~Sani \emph{et~al.}, ``Multivariable risk modelling and survival analysis with machine learning in sars-cov-2 infection,'' \emph{Journal of Clinical Medicine}, vol.~12, no.~22, p. 7164, 2023.

\bibitem{yang2022large_gatortron}
X.~Yang, A.~Chen, N.~PourNejatian, H.~C. Shin, K.~E. Smith, C.~Parisien, C.~Compas, C.~Martin, A.~B. Costa, M.~G. Flores \emph{et~al.}, ``A large language model for electronic health records,'' \emph{NPJ Digital Medicine}, vol.~5, no.~1, p. 194, 2022.

\bibitem{papanikolaou2020dare_gpt2}
Y.~Papanikolaou and A.~Pierleoni, ``Dare: Data augmented relation extraction with gpt-2,'' \emph{arXiv preprint arXiv:2004.13845}, 2020.

\bibitem{zhang2023intra}
Z.~Zhang, X.~Zhang, K.~Ichiji, I.~Bukovsk{\`y}, and N.~Homma, ``How intra-source imbalanced datasets impact the performance of deep learning for covid-19 diagnosis using chest x-ray images,'' \emph{Scientific Reports}, vol.~13, no.~1, p. 19049, 2023.

\bibitem{shin2016learning}
H.-C. Shin, K.~Roberts, L.~Lu, D.~Demner-Fushman, J.~Yao, and R.~M. Summers, ``Learning to read chest x-rays: Recurrent neural cascade model for automated image annotation,'' in \emph{Proceedings of the IEEE conference on computer vision and pattern recognition}, 2016, pp. 2497--2506.

\bibitem{xue2018multimodal}
Y.~Xue, T.~Xu, L.~Rodney~Long, Z.~Xue, S.~Antani, G.~R. Thoma, and X.~Huang, ``Multimodal recurrent model with attention for automated radiology report generation,'' in \emph{Medical Image Computing and Computer Assisted Intervention--MICCAI 2018: 21st International Conference, Granada, Spain, September 16-20, 2018, Proceedings, Part I}.\hskip 1em plus 0.5em minus 0.4em\relax Springer, 2018, pp. 457--466.

\bibitem{yuan2019automatic}
J.~Yuan, H.~Liao, R.~Luo, and J.~Luo, ``Automatic radiology report generation based on multi-view image fusion and medical concept enrichment,'' in \emph{Medical Image Computing and Computer Assisted Intervention--MICCAI 2019: 22nd International Conference, Shenzhen, China, October 13--17, 2019, Proceedings, Part VI 22}.\hskip 1em plus 0.5em minus 0.4em\relax Springer, 2019, pp. 721--729.

\bibitem{geis2019ethics}
J.~R. Geis, A.~P. Brady, C.~C. Wu, J.~Spencer, E.~Ranschaert, J.~L. Jaremko, S.~G. Langer, A.~Borondy~Kitts, J.~Birch, W.~F. Shields \emph{et~al.}, ``Ethics of artificial intelligence in radiology: summary of the joint european and north american multisociety statement,'' \emph{Radiology}, vol. 293, no.~2, pp. 436--440, 2019.

\bibitem{guidotti2018survey}
R.~Guidotti, A.~Monreale, S.~Ruggieri, F.~Turini, F.~Giannotti, and D.~Pedreschi, ``A survey of methods for explaining black box models,'' \emph{ACM computing surveys (CSUR)}, vol.~51, no.~5, pp. 1--42, 2018.

\bibitem{pahadia2020radiology}
M.~Pahadia, S.~Khurana, H.~Geha, and S.~T.~I. Deahl, ``Radiology report writing skills: A linguistic and technical guide for early-career oral and maxillofacial radiologists,'' \emph{Imaging Science in Dentistry}, vol.~50, no.~3, p. 269, 2020.

\bibitem{rsf}
H.~Ishwaran, U.~B. Kogalur, E.~H. Blackstone, and M.~S. Lauer, ``Random survival forests,'' 2008.

\bibitem{coxmodel}
J.~Fox and S.~Weisberg, ``Cox proportional-hazards regression for survival data,'' \emph{An R and S-PLUS companion to applied regression}, vol. 2002, 2002.

\bibitem{deepsurv}
J.~L. Katzman, U.~Shaham, A.~Cloninger, J.~Bates, T.~Jiang, and Y.~Kluger, ``Deepsurv: personalized treatment recommender system using a cox proportional hazards deep neural network,'' \emph{BMC medical research methodology}, vol.~18, no.~1, pp. 1--12, 2018.

\bibitem{filev2022structured}
P.~D. Filev and A.~E. Stillman, ``Structured reporting in medical imaging: The role of artificial intelligence,'' in \emph{Artificial Intelligence in Cardiothoracic Imaging}.\hskip 1em plus 0.5em minus 0.4em\relax Springer, 2022, pp. 105--112.

\bibitem{nicolson2023improving_cvt2}
A.~Nicolson, J.~Dowling, and B.~Koopman, ``Improving chest x-ray report generation by leveraging warm starting,'' \emph{Artificial intelligence in medicine}, vol. 144, p. 102633, 2023.

\bibitem{seyyed2021underdiagnosis}
L.~Seyyed-Kalantari, H.~Zhang, M.~B. McDermott, I.~Y. Chen, and M.~Ghassemi, ``Underdiagnosis bias of artificial intelligence algorithms applied to chest radiographs in under-served patient populations,'' \emph{Nature medicine}, vol.~27, no.~12, pp. 2176--2182, 2021.

\bibitem{chen2020generating_r2gen}
Z.~Chen, Y.~Song, T.-H. Chang, and X.~Wan, ``Generating radiology reports via memory-driven transformer,'' \emph{arXiv preprint arXiv:2010.16056}, 2020.

\bibitem{chen2022cross_cmn}
Z.~Chen, Y.~Shen, Y.~Song, and X.~Wan, ``Cross-modal memory networks for radiology report generation,'' \emph{arXiv preprint arXiv:2204.13258}, 2022.

\bibitem{jing2018automatic}
B.~Jing, P.~Xie, and E.~Xing, ``On the automatic generation of medical imaging reports,'' in \emph{Proceedings of the 56th Annual Meeting of the Association for Computational Linguistics (Volume 1: Long Papers)}, 2018, pp. 2577--2586.

\bibitem{yin2019automatic}
C.~Yin, B.~Qian, J.~Wei, X.~Li, X.~Zhang, Y.~Li, and Q.~Zheng, ``Automatic generation of medical imaging diagnostic report with hierarchical recurrent neural network,'' in \emph{2019 IEEE international conference on data mining (ICDM)}.\hskip 1em plus 0.5em minus 0.4em\relax IEEE, 2019, pp. 728--737.

\bibitem{you2021aligntransformer}
D.~You, F.~Liu, S.~Ge, X.~Xie, J.~Zhang, and X.~Wu, ``Aligntransformer: Hierarchical alignment of visual regions and disease tags for medical report generation,'' in \emph{Medical Image Computing and Computer Assisted Intervention--MICCAI 2021: 24th International Conference, Strasbourg, France, September 27--October 1, 2021, Proceedings, Part III 24}.\hskip 1em plus 0.5em minus 0.4em\relax Springer, 2021, pp. 72--82.

\bibitem{liu2021exploring_PPKED}
F.~Liu, X.~Wu, S.~Ge, W.~Fan, and Y.~Zou, ``Exploring and distilling posterior and prior knowledge for radiology report generation,'' in \emph{Proceedings of the IEEE/CVF conference on computer vision and pattern recognition}, 2021, pp. 13\,753--13\,762.

\bibitem{wang2021self}
Z.~Wang, L.~Zhou, L.~Wang, and X.~Li, ``A self-boosting framework for automated radiographic report generation,'' in \emph{Proceedings of the IEEE/CVF Conference on Computer Vision and Pattern Recognition}, 2021, pp. 2433--2442.

\bibitem{tanida2023interactive_rgrg}
T.~Tanida, P.~M{\"u}ller, G.~Kaissis, and D.~Rueckert, ``Interactive and explainable region-guided radiology report generation,'' in \emph{Proceedings of the IEEE/CVF Conference on Computer Vision and Pattern Recognition}, 2023, pp. 7433--7442.

\bibitem{ren2015faster}
S.~Ren, K.~He, R.~Girshick, and J.~Sun, ``Faster r-cnn: Towards real-time object detection with region proposal networks,'' \emph{Advances in neural information processing systems}, vol.~28, 2015.

\bibitem{wu2021chest}
J.~Wu, N.~Agu, I.~Lourentzou, A.~Sharma, J.~A. Paguio, J.~S. Yao, E.~C. Dee, S.~Kashyap, A.~Giovannini, L.~A. Celi \emph{et~al.}, ``Chest imagenome dataset for clinical reasoning,'' in \emph{Annual Conference on Neural Information Processing Systems}, 2021.

\bibitem{he2022masked}
K.~He, X.~Chen, S.~Xie, Y.~Li, P.~Doll{\'a}r, and R.~Girshick, ``Masked autoencoders are scalable vision learners,'' in \emph{Proceedings of the IEEE/CVF conference on computer vision and pattern recognition}, 2022, pp. 16\,000--16\,009.

\bibitem{cheng2023prior}
P.~Cheng, L.~Lin, J.~Lyu, Y.~Huang, W.~Luo, and X.~Tang, ``Prior: Prototype representation joint learning from medical images and reports,'' in \emph{Proceedings of the IEEE/CVF International Conference on Computer Vision}, 2023, pp. 21\,361--21\,371.

\bibitem{vaswani2017attention}
A.~Vaswani, N.~Shazeer, N.~Parmar, J.~Uszkoreit, L.~Jones, A.~N. Gomez, {\L}.~Kaiser, and I.~Polosukhin, ``Attention is all you need,'' \emph{Advances in neural information processing systems}, vol.~30, 2017.

\bibitem{alfarghaly2021automated}
O.~Alfarghaly, R.~Khaled, A.~Elkorany, M.~Helal, and A.~Fahmy, ``Automated radiology report generation using conditioned transformers,'' \emph{Informatics in Medicine Unlocked}, vol.~24, p. 100557, 2021.

\bibitem{ziegler2019encoder}
Z.~M. Ziegler, L.~Melas-Kyriazi, S.~Gehrmann, and A.~M. Rush, ``Encoder-agnostic adaptation for conditional language generation,'' \emph{arXiv preprint arXiv:1908.06938}, 2019.

\bibitem{johnson2019mimic}
A.~E. Johnson, T.~J. Pollard, S.~J. Berkowitz, N.~R. Greenbaum, M.~P. Lungren, C.-y. Deng, R.~G. Mark, and S.~Horng, ``Mimic-cxr, a de-identified publicly available database of chest radiographs with free-text reports,'' \emph{Scientific data}, vol.~6, no.~1, p. 317, 2019.

\bibitem{jiao2021prognostication}
Z.~Jiao, J.~W. Choi, K.~Halsey, T.~M.~L. Tran, B.~Hsieh, D.~Wang, F.~Eweje, R.~Wang, K.~Chang, J.~Wu \emph{et~al.}, ``Prognostication of patients with covid-19 using artificial intelligence based on chest x-rays and clinical data: a retrospective study,'' \emph{The Lancet Digital Health}, vol.~3, no.~5, pp. e286--e294, 2021.

\bibitem{papineni2002bleu}
K.~Papineni, S.~Roukos, T.~Ward, and W.-J. Zhu, ``Bleu: a method for automatic evaluation of machine translation,'' in \emph{Proceedings of the 40th annual meeting of the Association for Computational Linguistics}, 2002, pp. 311--318.

\bibitem{banerjee2005meteor}
S.~Banerjee and A.~Lavie, ``Meteor: An automatic metric for mt evaluation with improved correlation with human judgments,'' in \emph{Proceedings of the acl workshop on intrinsic and extrinsic evaluation measures for machine translation and/or summarization}, 2005, pp. 65--72.

\bibitem{chin2004rouge}
L.~Chin-Yew, ``Rouge: A package for automatic evaluation of summaries,'' in \emph{Proceedings of the Workshop on Text Summarization Branches Out, 2004}, 2004.

\bibitem{vedantam2015cider}
R.~Vedantam, C.~Lawrence~Zitnick, and D.~Parikh, ``Cider: Consensus-based image description evaluation,'' in \emph{Proceedings of the IEEE conference on computer vision and pattern recognition}, 2015, pp. 4566--4575.

\bibitem{liu2019clinically}
G.~Liu, T.-M.~H. Hsu, M.~McDermott, W.~Boag, W.-H. Weng, P.~Szolovits, and M.~Ghassemi, ``Clinically accurate chest x-ray report generation,'' in \emph{Machine Learning for Healthcare Conference}.\hskip 1em plus 0.5em minus 0.4em\relax PMLR, 2019, pp. 249--269.

\bibitem{irvin2019chexpert}
J.~Irvin, P.~Rajpurkar, M.~Ko, Y.~Yu, S.~Ciurea-Ilcus, C.~Chute, H.~Marklund, B.~Haghgoo, R.~Ball, K.~Shpanskaya \emph{et~al.}, ``Chexpert: A large chest radiograph dataset with uncertainty labels and expert comparison,'' in \emph{Proceedings of the AAAI conference on artificial intelligence}, vol.~33, no.~01, 2019, pp. 590--597.

\bibitem{cindex}
F.~E. Harrell~Jr, K.~L. Lee, R.~M. Califf, D.~B. Pryor, and R.~A. Rosati, ``Regression modelling strategies for improved prognostic prediction,'' \emph{Statistics in medicine}, vol.~3, no.~2, pp. 143--152, 1984.

\bibitem{selvaraju2017grad}
R.~R. Selvaraju, M.~Cogswell, A.~Das, R.~Vedantam, D.~Parikh, and D.~Batra, ``Grad-cam: Visual explanations from deep networks via gradient-based localization,'' in \emph{Proceedings of the IEEE international conference on computer vision}, 2017, pp. 618--626.

\end{thebibliography}

\end{document}